\documentclass[a4paper, 10 pt, onecolumn]{article}  % Comment this line out
\usepackage{a4wide}
\usepackage{tikz}
\usepackage{verbatim}
\usetikzlibrary{arrows}

\usetikzlibrary{%
    decorations.pathreplacing,%
    decorations.pathmorphing%
}

\usetikzlibrary{calc,patterns,
                 decorations.pathmorphing,
                 decorations.markings}

\usetikzlibrary{decorations.pathmorphing} % for snake lines
\usetikzlibrary{matrix} % for block alignment
\usetikzlibrary{arrows} % for arrow heads
\usetikzlibrary{calc} % for manimulation of coordinates

\tikzstyle{block} = [draw,rectangle,thick,minimum height=2em,minimum width=2em]
\tikzstyle{sum} = [draw,circle,inner sep=0mm,minimum size=2mm]
\tikzstyle{connector} = [->,thick]
\tikzstyle{line} = [thick]
\tikzstyle{branch} = [circle,inner sep=0pt,minimum size=1mm,fill=black,draw=black]
\tikzstyle{guide} = []
\tikzstyle{snakeline} = [connector, decorate, decoration={pre length=0.2cm,
                         post length=0.2cm, snake, amplitude=.4mm,
                         segment length=2mm},thick, magenta, ->]

\usepackage{amsmath,amssymb}
\usepackage{graphicx}
\usepackage{color}
\usepackage{indentfirst} %%%Ê׶οոñ

\usepackage{booktabs}%%%excel biaoge ºê
\usepackage{multirow}

\usepackage{subfigure}

\newtheorem{assumption}{Assumption}
\newtheorem{theorem}{Theorem}
\newtheorem{remark}{Remark}

\newtheorem{lemma}{Lemma}

\newtheorem{problem}{Problem}

\title{
Guaranteed Cost Tracking for Uncertain Coupled Multi-agent Systems
  Using Consensus over a Directed Graph\thanks{This work was supported by the Australian Research Council under projects DP0987369 and DP120102152.}
 \\[.5cm]
\normalsize To be presented at the 2013 Australian Control Conference,
Perth, Australia
}

\author{Yi Cheng\thanks{School of
    Engineering and Information Technology, The University of New South
    Wales at  the Australian Defence Force Academy, Canberra, ACT 2600,
    Australia. Email: {\tt\small y.cheng@adfa.edu.au,
      v.ougrinovski@adfa.edu.au}.}
\and V. Ugrinovskii%
\addtocounter{footnote}{-1}\footnotemark~%
\thanks{The work of V. Ugrinovskii was carried in part while he was a visitor at the Australian National University.}% <-this % stops a space
\and Guanghui Wen\thanks{Department of Mathematics, Southeast University,
  Nanjing 210096, China. Email: {\tt\small wenguanghui@gmail.com}.}
}

\begin{document}

\maketitle
\thispagestyle{empty}
\pagestyle{empty}

%%%%%%%%%%%%%%%%%%%%%%%%%%%%%%%%%%%%%%%%%%%%%%%%%%%%%%%%%%%%%%%%%%%%%%%%%%%%%%%%
\begin{abstract}

This paper considers the leader-follower control problem for a linear
multi-agent system with directed communication topology and linear nonidentical uncertain coupling
subject to integral quadratic constraints (IQCs). A consensus-type control
protocol is proposed based on each agent's states relative to its neighbors and leader's state relative to agents which observe the leader. A sufficient condition is obtained by overbounding the cost function. Based on this sufficient condition, a computational algorithm is introduced to minimize the proposed guaranteed bound on tracking performance, which yields a suboptimal bound on the system consensus control and tracking performance. The effectiveness of the proposed method is demonstrated using a simulation example.

\end{abstract}

%%%%%%%%%%%%%%%%%%%%%%%%%%%%%%%%%%%%%%%%%%%%%%%%%%%%%%%%%%%%%%%%%%%%%%%%%%%%%%%%
\section{Introduction}

In recent years, theoretical studies of distributed coordination and
control for multi-agent systems have attracted much attention in the literature, with broad applications in various areas including unmanned air vehicles (UAVs), formation control, flocking, distributed sensor networks, etc.~\cite{[Olfati2004]}. As a
result, much progress has been made in the study of cooperative control of
multi-agent systems \cite{[Olfati2007],[Ren2007],[A. Arenas]}.

Efforts have recently been made to consider the leader-following
consensus problem. For example, the leader-following consensus problem for higher
order multi-agent systems is presented for both fixed and switching
topologies in \cite{[Ni2010]}. In \cite{[Hong2008]}, distributed observers
are designed for the system of second-order agents where an active leader
to be followed moves with an unknown velocity, and the interaction topology
has a switching nature. The consensus-based approach to observer-based
synchronization of multi-agent systems to the leader has been explored
in~\cite{U6,U8}.

A common feature of the above literature on leader-following
consensus-based control problems is that interactions between agents are
not considered. However, in many physical systems, interactions between
agents are inevitable and must be taken into account. Examples of
systems with a dynamical interaction between subsystems include power
systems and spacecraft control systems \cite{[Siljak]}. This necessitates
considering systems of interconnected agents.

In this paper, the leader-follower control problem for multi-agent systems
coupled via linear unmodelled dynamics is considered. Coupling among the
agents is regarded as an uncertainty and is described in term of time
domain  integral quadratic constraints (IQCs) \cite{[Ian2000]}. The IQC
modeling is a well established technique to describe uncertain interactions
between subsystems in a large scale system
\cite{[Ugrinovskii2005],[Li2007],[Ugrinovskii2000]}.

The motivation of this paper is to extend our previous work
\cite{[Cheng2013]} as follows.
Firstly, this paper considers the multi-agent system with directed topology
rather than undirected topology, which poses additional difficulty compared
with \cite{[Cheng2013]}, due to the Laplacian matrix of
directed graphs being in general asymmetric. Therefore a different technique is used
in this paper to obtain a sufficient condition for leader follower tracking
which does not involve coordinate transformation; the latter was used in \cite{[Cheng2013]} and required the
Laplacian matrix of the graph to be symmetric. Furthermore, we consider a
more general, compared to \cite{[Cheng2013]}, class of systems with
nonidentical time varying uncertain coupling. In this paper, we also
propose a different LQR based cost function which describes the cost on the
tracking error between the leader and {\it all} of the followers. In
contrast, in \cite{[Cheng2013]} a consensus based cost function is
considered, which penalizes the system input, the state error between the
agent and its neighbours, as well as the tracking error between the leader
and {\it selected} agents which observe the leader. Furthermore, the
graph topology of the control protocol does not need to be the same as the
topology of interconnections between the agents. Even though both
communication topologies are represented as directed graphs, these graphs
can be different: the agents are coupled over one directed graph, but the
control protocol for the system uses another directed graph.

The main contribution of this paper is to propose a sufficient condition
for the design of a guaranteed performance leader-follower control protocol for
multi-agent systems with directed interconnection topology and a quite general
linear uncertain coupling subject to IQCs. The sufficient condition is
obtained by using a direct over-bounding technique and involves checking
feasibility of parameterized linear matrix inequalities (LMIs). The
computational algorithm is introduced to minimize the proposed guaranteed
bound by choosing local tuning parameters and guarantee a suboptimal bound
on the system tracking performance.

The remainder of the paper proceeds as follows. In Section
\ref{problem formulation} of the paper, we set up the leader
follower control problem for a multi-agent system with directed topology and
nonidentical linear uncertain coupling and give some preliminaries. The
main results are given in Section
\uppercase\expandafter{\romannumeral3}. In section
\uppercase\expandafter{\romannumeral4}, the computational algorithms are
introduced. Section \uppercase\expandafter{\romannumeral5} gives an example
which illustrates the theory presented in the paper. Finally, the
conclusions are given in Section \uppercase\expandafter{\romannumeral6}.

\section{Problem Formulation and Preliminaries}
\label{problem formulation}

\subsection{Graph theory}
Consider a directed graph $\mathcal {G}=(\mathcal {V}, \mathcal {E}, \mathcal {A})$, where $\mathcal {V}= \{1, 2,\cdots, N\}$ is a finite nonempty node set and $\mathcal {E} \subseteq \mathcal {V}\times \mathcal {V}$ is an edge set of ordered pairs of nodes. The edge $(i,j)$ in the edge set of an directed graph means that the node $i$  can obtain information from node $j$. Node $i$ is called a neighbor of node $j$ if $(i,j)\in \mathcal {E}$. The set of neighbors of node $i$ is defined as $N_i=\{j|(i,j)\in \mathcal {E}\}$. $\mathcal {G}$ is a simple graph if it has no self-loops or repeated edges. If there is a directed path between any two nodes of the graph $\mathcal {G}$, then the graph $\mathcal {G}$ is strongly connected. The adjacency matrix $\mathcal {A}=[a_{ij}]\in R^{N\times N}$ of the directed graph $\mathcal {G}$ is defined as $a_{ij}=1$ if $(i,j)\in \mathcal {E}$, and $a_{ij}=0$ otherwise. The in-degree matrix $\mathcal {D} =diag\{d_1,\cdots,d_N\}\in R^{N\times N}$ is a diagonal matrix, whose diagonal elements are $d_i = \sum\limits_{j=1}^{N}a_{ij}$ for $i = 1,\cdots,N$. Also, let $q_i = \sum\limits_{j=1}^{N}a_{ji}$ be the out-degree of node $i$. The Laplacian matrix of the graph is defined as
$
\mathcal {L} =\mathcal {D} - \mathcal {A}.
$

\subsection{Problem Formulation}

 Consider a system consisting of $N$ agents and a leader. All $N$ agents
 are assumed to be linear dynamical agents, coupled with their neighbors
 via, in general nonidentical, linear uncertain coupling. The connection
 between $N$ agents is described by a directed graph
 $\mathcal{G}_1$, with the
 node set $\mathcal {V}=\{1,\ldots, N\}$, an edge set $\mathcal {E}_1$
 and a corresponding adjacency matrix $\mathcal {A}_1$.
The dynamics of the $i$th agent are described as
\begin{equation} \label{agents dymamic}
 \dot{x}_i=Ax_i+B_1u_i+ B_2\sum\limits_{j\in S_i} \varphi_{ij} (t, x_j(.)|_0^t-x_i(.)|_0^t),
\end{equation}
where the summation is over the set $S_i$ of neighbors of node $i$ in the
graph $\mathcal{G}_1$. The notation $\varphi_{ij}(t, y (.)|_0^t)$ describes
a linear uncertain operator mapping functions $y(s)$, $0\leq s \leq t$ into
$\Re^m$. Also, $x_i\in \Re^n$ is the state, $u_i\in \Re^p$ is the control
input. We note that the last term in (\ref{agents dymamic}) reflects a
relative, time varying nature of interactions between agents.

Let $L_{2e}^n [0, \infty)$ be the space of functions $y(.): [0, \infty) \rightarrow \Re ^n$ such that $\int_{0}^{T}\|y(t)\|^2 dt < \infty,~\forall T >0$.

\begin{assumption}\label{A1}
All the mappings $\varphi_{ij}(., .)$
satisfy the following assumptions:
\begin{enumerate}
\item
$\forall y \in
L_{2e}^n[0, \infty), \varphi_{ij}(.,y(.)|_0^.) \in L_{2e}^m[0, \infty) $.
\item
$\forall t > 0, \varphi_{ij}(t, y)$ is linear in the second argument, i.e., if $y=\tau_1y_1+\tau_2y_2$, then $ \varphi_{ij}(t,y(.)|_0^t)=\tau_1 \varphi_{ij}(t,y_1(.)|_0^t)+\tau_2 \varphi_{ij}(t, y_2(.)|_0^t)$.
\item
Given a matrix $C_{ij}\in \Re^{r\times n}$, there exists a sequence $\{t_l\}, t_l \rightarrow \infty$, such that for every $t_l$, the following IQC holds
\begin{eqnarray} \label{IQC}
\int_{0}^{t_l}\|\varphi_{ij}(t,y(.)|_0^t)\|^2dt \leq \int_{0}^{t_l} \|C_{ij} y \|^2
dt, \\
\forall y \in L_{2e} [0, \infty). \nonumber
\end{eqnarray}
\end{enumerate}
The sequence $\{t_l\}$  is assumed to be the same for all $\varphi_{ij}$. The class of operators that satisfy these assumptions will be denoted by $\Xi_0$. We note that matrices $C_{ij}$ are assumed to be fixed.
\end{assumption}

%\begin{remark}\label{Rem1}
%This assumption captures some common classes of uncertain coupling. For
%example, $\varphi_{ij}$ can be a linear causal operator from the Hardy space
%$H^\infty$. Such operators have extension to operator mapping $L_{2e} [0,
%\infty)$ into $L_{2e} [0, \infty)$ \cite{[Willems1971]}. For instance, it
%is easy to show that unmodelled dynamics described as
%\begin{displaymath}
%\left\{
%\begin{array}{ll}
%\dot{\zeta}_i=-a_i{\zeta}_i+ y(t),\\
%\varphi_{ij} (t, y (.)|_0^t)=\zeta_i (t)
%\end{array}
%\right.
%\end{displaymath}
%satisfy (\ref{IQC}). Then the term $\varphi_{ij}(t, x_j(.)-x_i(.))$ in (\ref{agents
%  dymamic}) can be interpreted as an action based on relative measurements
%and applied through a dynamic channel with memory.
%
%Uncertain input delay in receiving relative states is also allowed. Let
%$\varphi_{ij}(t, y (.)|_0^t)= y(t-\tau)$. Then the corresponding coupling term
%\begin{displaymath}
%\varphi_{ij}(t, x_j(.)-x_i(.))= x_j(t-\tau) - x_i(t-\tau).
%\end{displaymath}
%Finally (\ref{IQC}) captures norm-bounded uncertain coupling
%$\varphi_{ij}(t, y(.)|_0^t)=\Delta y(t)$ where ${\Delta}$ is a constant matrix such
%that ${\Delta}^{'}\Delta \leq I$.
%\end{remark}

In addition to the system (\ref{agents dymamic}), suppose a leader is given. The dynamics of the leader, labeled $0$, is expressed as
\begin{equation} \label{leader dymamic}
 \dot{x}_0=Ax_0,
\end{equation}
where $x_0\in \Re^n$ is its state. The control communication topology
between $N$ agents is described by a directed graph $\mathcal {G}_2$,
with the same node set $\mathcal {V}=\{1,\ldots, N\}$, but possibly
different edge set $\mathcal {E}_2$ and a corresponding adjacency matrix
$\mathcal {A}_2$. The
Laplacian matrix of the graph $\mathcal {G}_2$ is denoted as $\mathcal
{L}_2$. We assume throughout the paper that the
leader node can be observed from a subset of nodes of the graph $\mathcal
{G}_2$. If the leader is observed by the node $i$, we extend the graph
$\mathcal{G}_2$ by adding the edge $(0,i)$ with weighting gain $g_i=1$,
otherwise let $g_i=0$. We refer to node $i$ with $g_i\neq 0$ as a pinned or
controlled node. The diagonal matrix $G=diag\{g_i\}\in \Re
^{N\times N}$ is commonly referred to as the pinning matrix. The system is
assumed to have at least one agent connected to the leader, hence $G \neq 0$.

Define error vectors as
$e_i=x_0-x_i$, $i=1,2,\ldots,N$.
Then dynamics of the synchronization errors satisfy the equation
\begin{equation} \label{error dymamic}
 \dot{e}_i=Ae_i-B_1u_i-B_2 \sum\limits_{j\in S_i} \varphi_{ij} (t, e_i(.)|_0^t-e_j(.)|_0^t).
\end{equation}

In this paper we are concerned with finding a control protocol for each
node $i$ of the form
\begin{equation} \label{controller}
 u_i=-K\{\sum\limits_{j\in T_i}(x_j-x_i) + g_i(x_0-x_i)\},
\end{equation}
where $K$ is the feedback gain matrix to be found, and $T_i$ is the set of
neighbors of node $i$ in the graph $\mathcal{G}_2$. As a measure of system
performance, we will use the quadratic cost function,
%\begin{align} \label{cost function}
% \mathcal {J}(u) &=\sum\limits_{i=1}^{N} \int_{0}^{\infty} \Big(\big( \frac{1}{w_i}(\sum \limits_{j \in S_i}(e_i-e_j)+g_i e_i\big)' Q_i \big( \frac{1}{w_i}(\sum \limits_{j \in S_i}(e_i-e_j)+g_i e_i\big) + u_i'R_i u_i\Big) dt
%\end{align}
%where $Q_i=Q_i'>0$ and $R=R'>0$ are given weighting matrices, $w_i=\sum\limits_{j=1}^{N} $ and $u$ denotes
%the vector $u=[u_1'~\ldots~u_N']'$.
%
%Let $Q_i= w_i^2 Q$ and
%\begin{align} \label{cost function 11}
% \mathcal {J}(u) = \int_{0}^{\infty} e'\Big((\mathcal{L}+G)'(\mathcal{L}+G) \otimes Q  +  (\mathcal{L}+G)'(\mathcal{L}+G) \otimes K' R K \Big)e dt,
%\end{align}
%
%Let $M=(\mathcal{L}+G)'(\mathcal{L}+G)$ and $M$ is a positive definite and symmetric matrix. Let $T\in \Re^{N\times N}$ be an orthogonal matrix such that
%\begin{align} \label{transformation}
%T^{-1}M T=J=\mathrm{diag}\left[\lambda_1, \cdots, \lambda_N\right].
%\end{align}
\begin{align} \label{cost function}
 \mathcal {J}(u) &=\sum\limits_{i=1}^{N} \int_{0}^{\infty} \Big(e_i'Q  e_i + u_i'R u_i\Big) dt,
\end{align}
where $Q=Q'>0$ and $R=R'>0$ are given weighting matrices.

\begin{remark}
In \cite{[Cheng2013]} we considered a different cost function,
\begin{eqnarray*}
\mathcal {J}'(u)= \sum\limits_{i=1}^{N}
\int_{0}^{\infty}(\frac{1}{2}\sum\limits_{j\in N_i}(e_i - e_j)' Q (e_i -
e_j) \\ + g_i e_i' Q e_i +u_i'R u_i)dt.
\end{eqnarray*}
Each addend in this cost function penalizes the $i$th system input, the disagreement between the $i$th and the $j$th system states, where $j$ is a neighbor of $i$, as well as
the tracking errors between the leader and the pinned agents which observe
the leader. In contrast, the cost function (\ref{cost function}) in this paper
describes the cost on the tracking error between the leader and {\it all}
of the followers and system input. \hfill$\Box$
\end{remark}

Taking linearity of the operator $\varphi_{ij}$ into account, the synchronization error dynamic (\ref{error dymamic}) can be represented as
\begin{equation} \label{error dymamic with norm}
 \dot{e}_i=Ae_i-B_1u_i - B_2 \sum\limits_{j\in S_i} (\varphi_{ij}(t, e_i(.)|_0^t) - \varphi_{ij}(t, e_j(.)|_0^t)).
\end{equation}

The problem in this paper is to find a control protocol (\ref{controller}) which solves the leader following consensus control problem as follows:
\begin{problem}\label{prob1}
Under Assumption~\ref{A1}, find a control protocol of the form
(\ref{controller}) such that
\begin{equation} \label{cost function 1}
\sup \limits_{\Xi_0}\mathcal {J}(u) < \infty.
\end{equation}
\end{problem}

Here $\sup \limits_{\Xi_0}$ means that the supremum is taken over the set of all operators $\varphi_{ij}$ that belong to the class $\Xi_0$ of operators.
Since $Q>0$, then (\ref{cost function 1}) implies
 \begin{equation} \label{synchronization}
\int_{0}^{\infty}\|e_i\|^2 dt
 < \infty \quad \forall i=1,\ldots,N.
\end{equation}
Hence, solving Problem~\ref{prob1} implies synchronization of all agents to the
leader in the $L_2$ sense.

%\subsection{Associated Guaranteed Cost Decentralized Control Problem}

\section{The Main Result}

In this section, the main result of this
paper is presented which is a sufficient condition for the system (1) to be
able to track the leader with guaranteed tracking
performance.

First we present the following result of \cite{Zhang2012} and some notation.

\begin{assumption}
\label{graph assumption}
 The digraph $\mathcal{G}_2$ contains a spanning tree and
the root node $i_r$ obtains information from the leader node, i.e., $g_{i_r}
> 0$.
\end{assumption}

\begin{lemma}\label{Zhang2012}(\cite{Zhang2012})
Under Assumption \ref{graph assumption}, $\mathcal{L}_2+G$ is nonsingular. Define $\left[\vartheta_1, \cdots, \vartheta_N\right]'= (\mathcal{L}_2+G)^{-1} 1_N$, $ \Theta=\mathrm{diag}\{\vartheta_i^{-1}\}$ and $H=\Theta(\mathcal{L}_2+G)+(\mathcal{L}_2+G)'\Theta$, then $\Theta>0$ and $H>0$.
\end{lemma}

Let $\sigma'$ be the maximum eigenvalue of $H$ and
$\sigma=\frac{1}{2}\sigma'$. Also let
$M=(\mathcal{L}_2+G)'(\mathcal{L}_2+G)$. According to
Lemma~\ref{Zhang2012}, $M$ is a positive definite and symmetric matrix. Let
$\mathcal{T}\in \Re^{N\times N}$ be an orthogonal matrix such that
\begin{align} \label{transformation}
\mathcal{T}^{-1}M \mathcal{T}=J=\mathrm{diag}\left[\lambda_1, \cdots, \lambda_N\right],
\end{align}
and denote $\bar \lambda= \max \limits_{i}(\lambda_i)$.
For node $i$ of the graph $\mathcal{G}_2$, introduce matrices $\hat
C_i=[C_{ij_1}' \ldots
C_{ij_{d_i}}']'$, $\bar C_i=[C_{r_1i}' \ldots
C_{r_{q_i}i}']'$, where $j_1,\ldots, j_{d_i}$ are the elements of the
neighbourhood  set $S_i$, and $r_1,\ldots,r_{q_i}$ are the nodes with the
property $(r_s,i)\in\mathcal{E}_1$; $d_i$ and $q_i$ are the in-degree and
the out-degree of node $i$, respectively, in the graph
$\mathcal{G}_1$. Also, introduce the matrix $\hat R= (\bar\lambda/\sigma) R$.

\begin{theorem}
\label{Theorem 1}
Let a matrix $ Y= Y'> 0, Y \in \Re ^{n\times n}$, and constants $\nu_{ij}>0$,
$\mu_{ij}>0$, $j\in S_i$, $i=1,\ldots, N$, exist such that the
following LMIs are satisfied simultaneously
\begin{equation}
\label{LMI Dircted}
\left[
\begin{array}{ccccc}
Z_i     & YQ ^{1/2}  & Y\hat C_i'&  Y \bar C_i'\\
 Q^{1/2}Y  & -\frac{1}{\vartheta_i}I & 0 & 0 \\
\hat C_iY  & 0 & -\Phi_i & 0  \\
\bar C_i Y  & 0 & 0  &  -\Omega_i
\end{array}\right]<0,
\end{equation}
 where
\begin{align*}
Z_i=&AY+YA' - \sigma \vartheta_i B_1\hat R^{-1} B_1'\\
 &+ \vartheta_i\sum \limits_{j\in S_i}(\frac{1}{\nu_{ij}}+ \frac{1}{\mu_{ij}}) B_2B_2',\\
\Phi_i=&\mathrm{diag}[\frac{\vartheta_i}{\nu_{ij}}, j\in S_i],\\
\Omega_i=&\mathrm{diag}[\frac{\vartheta_i}{\mu_{ji}}, j: i\in S_j].
\end{align*}
Then the control protocol (\ref{controller}) with $K=- (\sigma/\bar
  \lambda)R^{-1}B_1'Y^{-1}$ solves Problem~\ref{prob1}. Furthermore, this
protocol guarantees the following performance bound
 \begin{equation} \label{cost function TH1}
 \sup_{\Xi_0}\mathcal {J}(u) \le\sum_{i=1}^{N}\vartheta_i^{-1} e_i'(0)Y^{-1}e_i(0).
\end{equation}
\end{theorem}

%\begin{proof}
\emph{Proof: }
Using the Schur complement, the LMIs (\ref{LMI Dircted}) can be transformed into the following Riccati inequality
\begin{align}
\label{ARI}
&AY+YA' - \sigma \vartheta_i B_1 \hat R^{-1} B_1' + \vartheta_i\sum \limits_{j\in S_i}(\frac{1}{\nu_{ij}}+ \frac{1}{\mu_{ij}}) B_2B_2' \nonumber \\
& +Y(\vartheta_i Q + \vartheta_i^{-1} (\sum \limits_{j\in S_i} \nu_{ij}C'_{ij}C_{ij} \nonumber \\
&+ \sum\limits_{j: i \in S_j}\mu_{ji}C'_{ji}C_{ji}))Y < 0.
\end{align}

 After pre- and post-multiplying (\ref{ARI}) by $Y^{-1}$ and multiplying (\ref{ARI}) by $\vartheta_i^{-1}$, then substituting $K=- (\sigma/\bar
  \lambda)R^{-1}B_1'Y^{-1}$ into the Riccati inequality (\ref{ARI}), we obtain
\begin{align}
\label{ARI in2}
&Y^{-1} (\vartheta_i^{-1}A+ \sigma B_1K) + (\vartheta_i^{-1} A+ \sigma B_1K)'Y^{-1} \nonumber\\
&+ \sigma K' \hat R K  + \sum\limits_{j\in S_i}(\frac{1}{\nu_{ij}}+ \frac{1}{\mu_{ij}})Y^{-1}B_2B_2'Y^{-1} + Q \nonumber \\
&+ \vartheta_i^{-2}(\sum \limits_{j\in S_i} \nu_{ij}C'_{ij}C_{ij} + \sum\limits_{j: i \in S_j}\mu_{ji}C'_{ji}C_{ji}) < 0.
\end{align}

Define $e=[e'_1, \cdots, e'_N]'$ and consider the following Lyapunov
function candidate for the subsystems (\ref{error dymamic with norm}):
\begin{equation}
V(e)= \sum \limits_{i=1}^N \vartheta_i^{-1} e_i' Y^{-1}e_i.
\end{equation}
Then
\begin{align}
\label{lyapunov equation11}
 \frac{d V(e)}{dt}&=\sum \limits_{i=1}^N 2 e_i' Y^{-1}\Big(\vartheta_i^{-1} Ae_i \nonumber\\
 &+ \vartheta_i^{-1} B_1K(\sum \limits_{j\in T_i}(e_i-e_j)+g_ie_i) \Big) \nonumber\\
& -2\sum \limits_{i=1}^N\vartheta_i^{-1}  \sum\limits_{j\in S_i} e_i' Y^{-1}B_2  \varphi_{ij} (t, e_i(.)|_0^t) \nonumber\\
&+ 2\sum \limits_{i=1}^N\vartheta_i^{-1}  \sum\limits_{j\in S_i}e_i' Y^{-1}B_2  \varphi_{ij} (t, e_j(.)|_0^t).
\end{align}
Note the following inequality:
\begin{align}
\label{one trick}
&\sum \limits_{i=1}^N 2 e_i' \vartheta_i^{-1} Y^{-1} B_1K(\sum \limits_{j\in T_i}(e_i-e_j)+g_ie_i) \nonumber\\
&=2e'(\Theta(\mathcal {L}_2+G)\otimes (Y^{-1}B_1K))e \nonumber\\
&=e'\big((\Theta(\mathcal {L}_2+G)+ (\mathcal {L}_2+G)'\Theta)\otimes (Y^{-1}B_1\hat R^{-1}B_1'Y^{-1})\big)e \nonumber\\
&= y'\big((\Theta(\mathcal {L}_2+G)+ (\mathcal {L}_2+G)'\Theta)\otimes I_p\big)y \nonumber\\
&\leq 2 \sigma y'\big(I_N\otimes I_p\big)y = 2\sigma e'\big(I_N \otimes Y^{-1}B_1 \hat R^{-1}B_1'Y^{-1}\big)e \nonumber\\
&=2\sum \limits_{i=1}^N \sigma e_i' Y^{-1}B_1 \hat R^{-1}B_1'Y^{-1} e_i,
\end{align}
where $y=(I_N \otimes \hat R^{-1/2}B_1'Y^{-1}) e$.

From (\ref{lyapunov equation11}) and (\ref{one trick}), one has
\begin{align}
\label{lyapunov equation112}
\frac{d V(e)}{dt}&\leq \sum \limits_{i=1}^N 2 e_i' Y^{-1}\Big(\vartheta_i^{-1} A + \sigma B_1K \Big)e_i \nonumber\\
&-2\sum \limits_{i=1}^N\vartheta_i^{-1} \sum\limits_{j\in S_i} e_i' Y^{-1}B_2  \varphi_{ij} (t, e_i(.)|_0^t)  \nonumber\\
&+ 2\sum \limits_{i=1}^N\vartheta_i^{-1}  \sum\limits_{j\in S_i}e_i' Y^{-1}B_2  \varphi_{ij} (t, e_j(.)|_0^t).
\end{align}
%&=\sum \limits_{i=1}^N 2 e_i' Y^{-1}\Big(\vartheta_i^{-1} A + \sigma B_1K \Big)e_i -2\sum \limits_{i=1}^N\vartheta_i^{-1} \sum\limits_{j\in S_i} e_i' Y^{-1}B_2  \varphi_{ij} (t, e_i(.)|_0^t)  \nonumber\\
%&+ 2\sum \limits_{i=1}^N\vartheta_i^{-1}  e_i' Y^{-1} L \eta_i ,
%\end{align}

%where  $L= B_2 [I, \cdots, I, I,\cdots, I]$, $\eta_i=[a_{i,1}\varphi_{ij}(t, e_1 (.)|_0^t)',\cdots, a_{i,(i-1)}\varphi_{ij}(t, e_{i-1} (.)|_0^t)', a_{i,(i+1)}\varphi_{ij}(t, e_{i+1} (.)|_0^t)', \cdots,a_{i,N}\varphi_{ij}(t, e_N (.)|_0^t)']'$.

Substituting the Riccati inequality (\ref{ARI in2}) into (\ref{lyapunov equation112}), we have

\begin{align}
\label{lyapunov equation33}
\frac{d V(e)}{dt} &\leq - \sum \limits_{i=1}^Ne_i' \Big( \sigma K' \hat R K+ Q  \nonumber\\
& +\sum\limits_{j\in S_i}(\frac{1}{\nu_{ij}}+ \frac{1}{\mu_{ij}})Y^{-1}B_2B_2'Y^{-1}  \nonumber\\
& + \vartheta_i^{-2} (\sum \limits_{j\in S_i} \nu_{ij}C'_{ij}C_{ij} + \sum\limits_{j: i \in S_j}\mu_{ji}C'_{ji}C_{ji})\Big)e_i  \nonumber\\
&-2\sum \limits_{i=1}^N \vartheta_i^{-1}\sum\limits_{j\in S_i} e_i' Y^{-1}B_2  \varphi_{ij} (t, e_i(.)|_0^t) \nonumber \\
&+ 2\sum \limits_{i=1}^N\vartheta_i^{-1}  \sum\limits_{j\in S_i}e_i' Y^{-1}B_2  \varphi_{ij} (t, e_j(.)|_0^t).
 %&= - \sum \limits_{i=1}^Ne_i' \Big( \sigma K' \hat R K +  Q \Big)e_i
%- \sum \limits_{i=1}^{N}\sum\limits_{j\in S_i}\frac{1}{\nu_{ij}} e_i' Y^{-1}B_2B_2'Y^{-1}e_i\nonumber \\
%& -2\sum \limits_{i=1}^N \vartheta_i^{-1}\sum\limits_{j\in S_i} e_i' Y^{-1}B_2  \varphi_{ij} (t, e_i(.)|_0^t)
%- \sum \limits_{i=1}^{N} \sum\limits_{j\in S_i} \nu_{ij}  \vartheta_i^{-2} \| \varphi_{ij} (t, e_i(.)|_0^t)\|^2 \nonumber \\
%&+ \sum \limits_{i=1}^{N} \sum\limits_{j\in S_i} \vartheta_i^{-2} \nu_{ij} \| \varphi_{ij} (t, e_i(.)|_0^t)\|^2 - \sum \limits_{i=1}^{N} \sum\limits_{j\in S_i}\vartheta_i^{-2} \nu_{ij} \|C_{ij} e_i\|^2 - \sum \limits_{i=1}^{N}\sum\limits_{j\in S_i}\frac{1}{\mu_{ij}} e_i' Y^{-1}B_2B_2'Y^{-1}e_i  \nonumber \\
%&+ 2\sum \limits_{i=1}^N\vartheta_i^{-1}  \sum\limits_{j\in S_i}e_i' Y^{-1}B_2  \varphi_{ij} (t, e_j(.)|_0^t) - \sum \limits_{i=1}^{N} \sum\limits_{j\in S_i} \vartheta_i^{-2} \mu_{ij} \| \varphi_{ij} (t, e_j(.)|_0^t)\|^2 \nonumber \\
%& + \sum \limits_{i=1}^{N} \sum\limits_{j\in S_i} \vartheta_i^{-2} \mu_{ij} \| \varphi_{ij} (t, e_j(.)|_0^t)\|^2 - \sum \limits_{i=1}^{N} \sum\limits_{j: i\in S_j}\vartheta_i^{-2} \mu_{ji} \|C_{ji}e_i\|^2.
\end{align}

Using the following identity,
%\begin{align*}
%&\sum\limits_{i=1}^{N} \sum\limits_{j\in S_i} \mu_{ij}e'_j C'_{ij}C_{ij}e_j =\sum\limits_{i=1}^{N} \sum\limits_{j=1}^{N} a_{ij} \mu_{ij}e'_j C'_{ij}C_{ij}e_j \\
%&=\sum\limits_{j=1}^{N} \sum\limits_{i=1}^{N} a_{ij} \mu_{ij}e'_j C'_{ij}C_{ij}e_j =\sum\limits_{i=1}^{N} \sum\limits_{j=1}^{N} a_{ji}\mu_{ji}e'_i C'_{ji}C_{ji}e_i\\
%&= \sum\limits_{i=1}^{N} \sum\limits_{j: i \in S_j} \mu_{ji}e'_i C'_{ji}C_{ji}e_i
%%\label{identity}
%\end{align*}
\begin{align*}
\sum\limits_{i=1}^{N} \sum\limits_{j\in S_i} \mu_{ij}e'_j C'_{ij}C_{ij}e_j = \sum\limits_{i=1}^{N} \sum\limits_{j: i \in S_j} \mu_{ji}e'_i C'_{ji}C_{ji}e_i,
%\label{identity}
\end{align*}
one has
\begin{align}
\label{lyapunov equation44}
&\frac{d V(e)}{dt} \leq  - \sum \limits_{i=1}^{N} e_i' \Big(\sigma K' \hat R K + Q \Big)e_i\nonumber \\
&-\sum \limits_{i=1}^{N} \sum\limits_{j\in S_i} \|\frac{1}{\sqrt {\nu_{ij}}}B_2'Y^{-1}e_i + {\sqrt {\nu_{ij}}}\vartheta_i^{-1} \varphi_{ij} (t, e_i(.)|_0^t)\|^2 \nonumber \\
&+ \sum \limits_{i=1}^{N}\sum\limits_{j\in S_i}\vartheta_i^{-2} \nu_{ij} (\|\varphi_{ij} (t, e_i(.)|_0^t)\|^2-\|C_{ij}e_i\|^2 )\nonumber \\
&- \sum \limits_{i=1}^{N}\sum\limits_{j\in S_i}\| \frac{1}{\sqrt {\mu_{ij}}}B_2'Y^{-1}e_i - {\sqrt {\mu_{ij}}}\vartheta_i^{-1} \varphi_{ij} (t, e_j(.)|_0^t)\|^2 \nonumber \\
&+ \sum \limits_{i=1}^{N}\sum\limits_{j\in S_i} \vartheta_i^{-2} \mu_{ij}(\| \varphi_{ij} (t, e_j(.)|_0^t)\|^2 - \|C_{ij}e_j\|^2).
\end{align}

According to the IQC condition (\ref{IQC}), we have
\begin{align}
\label{lyapunov equation441}
\int_{0}^{t_l}\frac{d V(e)}{dt}dt  \leq -  \sum \limits_{i=1}^{N}  \int_{0}^{t_l}e_i' \Big(\sigma K' \hat R K + Q \Big)e_i  dt.
\end{align}
Since $V (e(t_l)) \ge 0$, then (\ref{lyapunov equation441}) implies
\begin{align}
 \sum \limits_{i=1}^{N}  \int_{0}^{t_l}e_i' \Big(\sigma K' \hat R K + Q \Big)e_i  dt  \leq V(e(0)).
\end{align}

The expression on the right hand side of the above inequality is independent of $t_l$. Letting $t_l\rightarrow \infty$  leads to
\begin{align}
 \sum \limits_{i=1}^{N}  \int_{0}^\infty e_i' \Big(\sigma K' \hat R K + Q \Big)e_i  dt \leq V(e(0)).
\end{align}
Using (\ref{cost function}) and (\ref{controller}), we have
\begin{align}
 \mathcal {J}(u)&=\sum\limits_{i=1}^{N} \int_{0}^{\infty} \Big(e_i'Q  e_i + u_i'R u_i\Big) dt  \nonumber \\
&=\int_{0}^{\infty} \Big(e'(I_N \otimes Q ) e \nonumber\\
&+ e'[(\mathcal{L}_2+G)'(\mathcal{L}_2+G)\otimes K' R K]e\Big) dt \nonumber\\
%=&\int_{0}^{\infty}e'(I_N \otimes Q ) e dt  \nonumber\\
%  &+ \int_{0}^{\infty}e'[(L+G)'(L+G)\otimes K' R K]e dt  \nonumber\\
   &\leq \int_{0}^{\infty} \Big(e'(I_N \otimes Q ) e + e'[I_N\otimes \bar \lambda K' R K]e\Big) dt  \nonumber\\
   &=\sum \limits_{i=1}^{N}  \int_{0}^\infty e_i' \Big(\bar \lambda K' R K + Q \Big)e_i  dt.
\end{align}

Since $\hat R= \frac{\bar \lambda}{\sigma} R$, then we obtain
\begin{align}
\mathcal{J}(u) &\leq \sum \limits_{i=1}^{N}  \int_{0}^{\infty} e_i'\Big( \sigma K'\hat R K + Q  \Big)e_i  dt \nonumber\\
&\leq \sum_{i=1}^{N}\vartheta_i^{-1} e_i'(0)Y^{-1}e_i(0).
\end{align}
It implies that the control protocol (\ref{controller}) with $K=-(\sigma/\bar
  \lambda)R^{-1}B_1'Y^{-1}$ solves Problem~\ref{prob1}, and also guarantees
  the performance bound (\ref{cost function TH1}).
%\end{proof}
\hfill$\Box$

\section{The Computational Algorithm}\label{Algorithm}

In this section, we provide an algorithm to calculate a suboptimal control gain
$K$. According to Theorem~\ref{Theorem 1}, the upper bound on tracking
performance is given by the right hand side of (\ref{cost function
  TH1}). Hence, one can achieve a suboptimal guaranteed performance by
optimizing this upper bound over the feasibility set of the LMIs~(\ref{LMI Dircted}):
\begin{eqnarray} \label{LMIAlg}
\mathcal{J}^*_{\mbox{(\ref{LMI Dircted})}}
=\inf \sum_{i=1}^{N} \vartheta_i^{-1} e_i'(0)Y^{-1}e_i(0),
\end{eqnarray}
where the infimum is taken over the feasibility set of the LMIs (\ref{LMI
  Dircted}), $\{Y,\nu_{ij},\mu_{ij},i=1\ldots,N, j\in S_i:~ \mbox{(\ref{LMI
    Dircted}) holds}\}$.

As in \cite{[Li2007]}, the optimization problem (\ref{LMIAlg}) can be shown
to be equivalent to the minimization of $\gamma$ subject to the constraints
\begin{eqnarray}
\label{minimize cost}
\gamma > \sum_{i=1}^{N} \vartheta_i^{-1} e_i'(0)Y^{-1}e_i(0), ~i=1,\cdots,N.
\end{eqnarray}
By the Schur complement, (\ref{minimize cost}) is equivalent to the LMI
\begin{align}
\label{LMI minimize
      cost}
\left[\begin{array}{cccccc}
    \gamma               &   e'(0)\\
    e(0)                 &  \Upsilon 
\end{array}\right]>0, ~i=1,\cdots,N,
\end{align}
where  
\begin{align*}
e(0)=&[e_1(0)'~ e_2(0)'~\ldots~e_N(0)']', \\
\Upsilon=&\mathrm{diag}[\vartheta_i Y, i=1,2,\cdots,N].
\end{align*}

This leads us to introduce the following optimization problem in the
variables $\gamma, Y, \frac{1}{\nu_{ij}}$ and $\frac{1}{\mu_{ij}}$: Find
\begin{equation}
\label{LMI}
\mathcal{J}_{\mbox{(\ref{LMI Dircted}),(\ref{LMI minimize
      cost})}}^{*}\triangleq \inf \gamma,
\end{equation}
where the infimum is with respect to $\gamma, Y, \frac{1}{\nu_{ij}}$ and $\frac{1}{\mu_{ij}}$ subject to (\ref{LMI Dircted}) and (\ref{LMI minimize cost}).

We conclude this discussion by stating equivalence between the
optimization problems (\ref{LMIAlg}) and (\ref{LMI}).
\begin{theorem}
\label{Theorem 3}
$\mathcal{J}_{\mbox{(\ref{LMI Dircted})}}^{*}
= \mathcal{J}_{\mbox{(\ref{LMI Dircted}),(\ref{LMI minimize
      cost})}}^{*}$.
\end{theorem}

%\begin{proof}
\emph{Proof: }
The proof of this theorem is similar to the proof of Theorem 15 in \cite{[Li2007]}.
%\end{proof}
\hfill$\Box$

Based on the foregoing discussion, we propose an algorithm for the design of the
suboptimal protocol (\ref{controller}) based on Theorems~\ref{Theorem 1}
and~\ref{Theorem 3}:
\begin{itemize}
\item
Solve the optimization problem (\ref{LMI}), to a desired
accuracy, obtaining a feasible collection $Y$, $\frac{1}{\nu_{ij}}$,
$\frac{1}{\mu_{ij}}$ and $\gamma$. The collection $(Y,
\frac{1}{\nu_{ij}}, \frac{1}{\mu_{ij}})$ then belongs to the feasibility set of
the LMIs (\ref{LMI Dircted}).
\item
Using the found matrix $Y$, construct the gain matrix $K$ to be used in
(\ref{controller}), by letting $K=- (\sigma/\bar
  \lambda)R^{-1}B_1'Y^{-1}$.
Also, the guaranteed consensus performance bound for
this protocol can be computed, using the expression on the right-hand side
of equation (\ref{cost function TH1}).
\end{itemize}

It must be noted that the above algorithms require the knowledge of initial
conditions of the leader and agents. In practice, however, the initial state
of the leader may not be known. It is possible to avoid using
$e_i(0)$ in these algorithms using the approaches outlined in~\cite{[Li2007]}.

\section{Example}
\begin{figure}
\centering
\begin{tikzpicture}[scale=1.1,
    media/.style={font={\footnotesize\sffamily}},
    wave/.style={
        decorate,decoration={snake,post length=1.4mm,amplitude=2mm,
        segment length=2mm},thick},
    interface/.style={
        postaction={draw,decorate,decoration={border,angle=-45,
                    amplitude=0.3cm,segment length=2mm}}},
    gluon/.style={decorate, draw=black,
        decoration={coil,amplitude=4pt, segment length=5pt}},scale=0.7
    ]

    \tikzstyle{damper}=[thick,decoration={markings,
   mark connection node=dmp,
   mark=at position 0.5 with
   {
     \node (dmp) [thick,inner sep=0pt,transform shape,rotate=-90,minimum
 width=15pt,minimum height=3pt,draw=none] {};
     \draw [thick] ($(dmp.north east)+(2pt,0)$) -- (dmp.south east) -- (dmp.south
 west) -- ($(dmp.north west)+(2pt,0)$);
     \draw [thick] ($(dmp.north)+(0,-5pt)$) -- ($(dmp.north)+(0,5pt)$);
   }
 }, decorate]

    \draw[blue,line width=1pt](-5.2,0)--(4.2,0);
    % Vertical dashed line
    \draw[dashed,black](-5,-2.3)--(-5,0);
    \draw[dashed,black](-2,-2.3)--(-2,0);
    \draw[dashed,black](1,-2.3)--(1,0);
    \draw[dashed,black](4,-2.3)--(4,0);

    \draw[thick](0:-5cm)--(46:-5.7cm)node[midway]{\scriptsize $~~l$};
    \path (-5,0)++(-84:2cm)node{\scriptsize  $\alpha_0$};

    \draw[->] (-5,-1.5) arc (-90:-63:.75cm);
    \draw[thick](0:-2cm)--(76:-4.1cm)node[midway]{\scriptsize $~~a$};
    \path (-2,0)++(-84:2cm)node{\scriptsize  $\alpha_1$};

    \draw[->] (-2,-1.5) arc (-90:-63:.75cm);
    \draw[thick](0:1cm)--(116:-4.6cm)node[midway]{\scriptsize $~~b$};
    \path (1,0)++(-84:2cm)node{\scriptsize  $\alpha_2$};

    \draw[->] (1,-1.5) arc (-90:-63:.75cm);
    \draw[thick](0:4cm)--(141:-6.4cm);
    \path (4,0)++(-84:2cm)node{\scriptsize $\alpha_3$};

    \draw[->] (4,-1.5) arc (-90:-63:.75cm);

 %%%%%%Damper%%%%%%%

 \draw[damper] ($(-0.7,-3.6) + (0,0.5)$) -- ($(0.9,-3.6) + (0,0.5)$);

  \draw[damper] ($(2.3,-4.0) + (0,0.5)$) -- ($(3.9,-4.0) + (0,0.5)$);
 %%%%%%Spring%%%%%%%
  \draw[thick,gluon]
       (-0.7,-2.1)--(0.9,-2.1);

  \draw[thick,gluon]
        (2.3,-2.5)--(3.9,-2.5);

 \draw [thick] (-0.7,-2.1)--(-0.7,-3.12);
 \draw [thick] (0.9,-2.09)--(0.9,-3.12);
  \draw [thick] (-1.3,-2.6)--(-0.7,-2.6);
 \draw [thick] (0.9,-2.6)--(1.7,-2.6);

 \draw [thick] (2.3,-2.5)--(2.3,-3.50);
 \draw [thick] (3.9,-2.49)--(3.9,-3.50);
  \draw [thick] (1.7,-3.0)--(2.3,-3.0);
 \draw [thick] (3.9,-3.0)--(4.69,-3.0);

  \filldraw[fill=white,line width=1pt](-1.34,-2.6) circle(.025cm);
     \filldraw[fill=white,line width=1pt](1.65,-2.6)  circle(.025cm);
          \filldraw[fill=white,line width=1pt](1.74,-3.0)  circle(.025cm);
     \filldraw[fill=white,line width=1pt](4.73,-3.0)circle(.025cm);

    \filldraw[fill=white,line width=1pt](-4,-4)circle(.2cm)node[right]{\scriptsize $~Leader$};
       \filldraw[fill=white,line width=1pt](-1,-4)circle(.2cm)node[right]{\scriptsize $~\leftarrow u_1$};
          \filldraw[fill=white,line width=1pt](2,-4)circle(.2cm)node[left]{\scriptsize $u_2 \rightarrow~$};
             \filldraw[fill=white,line width=1pt](5,-4)circle(.2cm)node[left]{\scriptsize $ u_3\rightarrow~$};
\end{tikzpicture}
\caption{Interconnected pendulums.}
\label{pendulums}
\end{figure}
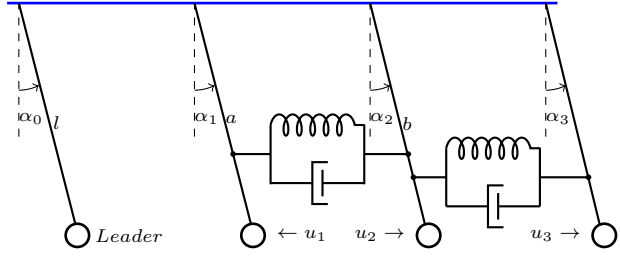

To illustrate the proposed method, consider a system consisting of three
identical pendulums coupled by two spring-damper systems. Each pendulum is subject to an
input as shown in Fig.~\ref{pendulums}. The dynamics of the coupled system
are governed by the following equations
\begin{align}
\label{dynamic of pedulums}
ml^2\ddot{\alpha}_1=&-k_{11}a^2(t)(\alpha_1-\alpha_2)-k_{12}a^2(t)(\dot\alpha_1-\dot\alpha_2)\nonumber\\
&-mgl\alpha_1-u_1,  \nonumber\\
ml^2\ddot{\alpha}_2=&-k_{21}b^2(t)(\alpha_2-\alpha_3)-k_{11}a^2(t)(\alpha_2-\alpha_1)\nonumber \\
 &-k_{22}b^2(t)(\dot\alpha_2-\dot\alpha_3)-k_{12}a^2(t)(\dot\alpha_2-\dot\alpha_1)\\
 &-mgl\alpha_2-u_2, \nonumber\\
ml^2\ddot{\alpha}_3=&-k_{21}b^2(t)(\alpha_3-\alpha_2)-k_{22}b^2(t)(\dot \alpha_3-\dot \alpha_2)\nonumber\\
&-mgl\alpha_3-u_3,\nonumber
\end{align}
where $l$ is the length of the pendulum, $a(t)$ and $b(t)$ are the
positions of the spring-damper, $g$ is the gravitational acceleration
constant, $m$ is the mass of each pendulum, $k_{11}$ and $k_{12}$ are the
spring constant and damping coefficient for the leftmost spring-damper  pair,
while $k_{21}$ and $k_{22}$ are the spring constant and damping coefficient
for the rightmost spring-damper pair. The position
of the spring-damper can change along the full length of the pendulums and
is considered to be uncertain, that is $0< a(t) \leq l$, $0< b(t) \leq l$.

In addition to the three pendulums, consider the leader pendulum which is
identical to those given. Its dynamics are described by the equation
\begin{equation}
\label{leader pedulums}
ml^2\ddot{\alpha}_0=-mgl\alpha_0.
\end{equation}

Choosing the state vectors as $x_0=(\alpha_0, \dot{\alpha}_0)$, $x_1=(\alpha_1, \dot{\alpha}_1)$, $x_2=( \alpha_2, \dot{\alpha}_2)$ and $x_3=(\alpha_3, \dot{\alpha}_3)$, equations (\ref{dynamic of pedulums}) and (\ref{leader pedulums}) can be written in the form of (\ref{agents dymamic}), (\ref{leader dymamic})
, where
$A=\left[\begin{array}{cc}
                          0  &  1  \\
                        -\frac{g}{l} &  0
                        \end{array}\right]$, $B_1=\left[\begin{array}{c}
                          0    \\
                        -\frac{1}{ml^2}
                        \end{array}\right]$, $B_2=\left[\begin{array}{cc}
                                0        \\
                         \frac{1}{m}
                        \end{array}\right]$ and $\varphi_{12}(t,x_2-x_1)=\frac{a^2(t)}{l^2} [k_{11} ~k_{12}] (x_2-x_1)$, $\varphi_{21}(t,x_1-x_2)=\frac{a^2(t)}{l^2} [k_{11} ~k_{12}] (x_1-x_2)$, $\varphi_{23}(t,x_3-x_2)=\frac{b^2(t)}{l^2} [k_{21} ~k_{22}] (x_3-x_2)$, $\varphi_{32}(t,x_2-x_3)=\frac{b^2(t)}{l^2} [k_{21} ~k_{12}] (x_2-x_3)$,

\begin{figure}
 \centering
 \begin{tikzpicture}[->,>=stealth',shorten >=1pt,auto,node distance=2cm,
   %thick,main node/.style={circle,fill=blue!20,draw,font=\sffamily\Large\bfseries}]
   thick,main node/.style={circle,fill=blue!20,draw,font=\sffamily\bfseries},]

  % \node[main node] (1) {0};
 %  \node[main node] (2) [below left of=1] {1};
 %  \node[main node] (3) [below right of=2] {3};
 %  %\node[main node] (4) [below left of=1] {3};
 %  \node[main node] (5) [below right of=1] {2};

   % \node[main node] (1) {0};
   \node[main node] (2)  {1};
   \node[main node] (3) [right of=2] {2};
   %\node[main node] (4) [below left of=1] {3};
   \node[main node] (5) [right of=3] {3};

   \path[every node/.style={font=\sffamily\small}]
     %(1) %edge node [left] {0.6} (4)
%        edge [right] node[left] {} (2)
%        %edge [bend right] node[left] {0.3} (2)
%         %edge [loop above] node {0.1} (1)
     (2) %edge node [right] {} (1)
       edge node {} (3)
        % edge [loop left] node {0.4} (2)
        % edge [right] node[left] {0.1} (3)%edge [bend right] node[left] {0.1} (3)
     (5) edge node [right] {} (3)
         %edge [right] node[right] {0.2} (4)%edge [bend right] node[right] {0.2} (4)
     %(4) edge node [left] {} (2)
         % edge [loop right] node {0.6} (4)
         % edge [bend right] node[right] {0.2} (1);
     (3) %edge node [right] {} (1)
        edge node {} (2)
        % edge [loop right] node {0.6} (4)
        edge [right] node[right] {} (5);  %edge [bend right] node[right] {0.2} (3);
 \end{tikzpicture}
 \centering
 \caption{The graph of uncertain interconnections between pendulums.}
 \label{coupling graph}
 \end{figure}
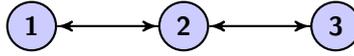

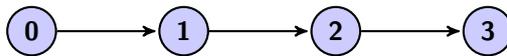
\begin{figure}
 \centering
 \begin{tikzpicture}[->,>=stealth',shorten >=1pt,auto,node distance=2cm,
   %thick,main node/.style={circle,fill=blue!20,draw,font=\sffamily\Large\bfseries}]
   thick,main node/.style={circle,fill=blue!20,draw,font=\sffamily\bfseries},]

  % \node[main node] (1) {0};
 %  \node[main node] (2) [below left of=1] {1};
 %  \node[main node] (3) [below right of=2] {3};
 %  %\node[main node] (4) [below left of=1] {3};
 %  \node[main node] (5) [below right of=1] {2};

    \node[main node] (1) {0};
   \node[main node] (2) [right of=1] {1};
   \node[main node] (3) [right of=2] {2};
   %\node[main node] (4) [below left of=1] {3};
   \node[main node] (4) [right of=3] {3};

   \path[every node/.style={font=\sffamily\small}]
     (1) %edge node [left] {0.6} (4)
        edge [right] node[left] {} (2)

     (2) %edge node [right] {} (1)
       edge node {} (3)

     (3) %edge node [right] {} (1)

        % edge [loop right] node {0.6} (4)
        edge [right] node[right] {} (4);%edge [bend right] node[right] {0.2} (3);

 \end{tikzpicture}
 \centering
 \caption{Directed communication graph for control.}
 \label{control graph}
 \end{figure}

The agents in this example are coupled according to the undirected graph
shown in Fig.~\ref{coupling graph}, which is treated here as a
special case of directed graph with symmetric adjacency matrix. On the
other hand, the control communication topology of the system is assumed to
be a linear directed graph shown in Fig.~\ref{control graph}. According to
this graph, only agent $1$ observes the leader. The Laplacian matrix
of the graph $\mathcal{G}_2$ consisting of nodes 1, 2 and 3 and the pinning
matrix $G$ are
\begin{displaymath}
\mathcal {L}_2=\left[\begin{array}{ccc}
                         0    &    0    &   0\\
                         -1   &    1    &   0\\
                         0    &   -1    &   1
                        \end{array}\right],~~
                        G=\left[\begin{array}{ccc}
                         1     &    0    &   0\\
                         0     &    0    &   0\\
                         0     &    0    &   0
                        \end{array}\right].
\end{displaymath}

\begin{figure}[htbp]
\centering
\includegraphics[width=.95\columnwidth]{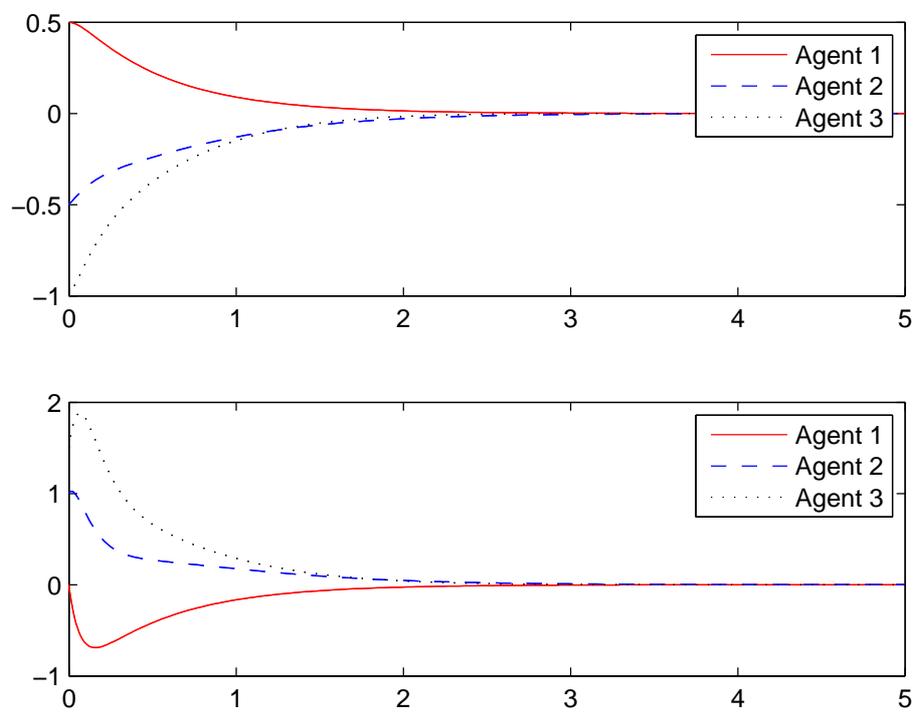}
\caption{Relative angles (the top figure) and relative velocities of the
  pendulums with respect to the leader, obtained using the algorithm based on
  Theorems~\ref{Theorem 1} and~\ref{Theorem 3}.}
\label{Theorem1Simu}
\end{figure}

To illustrate the design based on Theorems~\ref{Theorem 1} and
\ref{Theorem 3}, the LMI problem in Theorem \ref{Theorem 3} was solved numerically,
and then the trajectories of the coupled pendulum system with the obtained
protocol were simulated. To this end, the parameters of the coupled
pendulum system were chosen to be $m=0.25kg$, $l=1m$, $g=10m/s^2$, $
k_{11}=2 N/m$, $ k_{12}=1 N/(m/s)$, $ k_{21}=4 N/m$, $ k_{22}=2 N/(m/s)$,
$a=0.5\sin(0.2t)$, $b=0.8\cos(0.1t)$. In the cost function, we let
$Q=I$ and $R=0.1$. Using  the computational algorithm based on
Theorem~\ref{Theorem 3}, the
problem (\ref{LMIAlg}) was found to be feasible and yielded the gain matrix
$K=[4.5206, 4.2657]$. The performance bound was minimized by $\gamma=2.3532$, with parameters
$\nu_{12}=0.3268$, $\nu_{21}=0.6009$, $\nu_{23}=0.3007$, $\nu_{32}=0.1433$,
$\mu_{12}=0.1451$, $\mu_{21}=1.4300$, $\mu_{23}=0.1066$,
$\mu_{32}=0.4016$. The simulation results for this protocol are shown in
Fig.~\ref{Theorem1Simu}. Also, using the controller obtained by means of
the computational algorithm proposed in Theorem~\ref{Theorem 3}, we
directly computed the performance cost (\ref{cost function}) for the system
to be $\mathcal{J}(u)=2.3374$, while the theoretically predicted bound is
$\mathcal{J}_{\mbox{(\ref{LMI Dircted}),(\ref{LMI minimize cost})}}^{*}
= 2.3532$.
 
\section{Conclusions}
The consensus control for leader-tracking problem with guaranteed tracking
performance for nonidentical uncertain coupled linear systems connected over a directed graph has been discussed in this
paper. A sufficient condition was proposed by using the direct overbounding of the performance cost. According to the simulation results, the proposed computational algorithm based on Theorems 1, 2, which
solve N coupled LMIs, guarantees a suboptimal performance.

% that's all folks
\end{document}